\documentclass[12pt]{iopart}

\usepackage{graphics}
\usepackage{color}
\begin{document}

\title[Bottle-brush polymers with a flexible backbone under poor solvent
conditions]{Molecular dynamics simulations of single-component
bottle-brush polymers with a flexible backbone under poor solvent
conditions}

\author{Nikolaos G Fytas$^1$ and Panagiotis E Theodorakis$^2$}

\address{$^1$ Applied Mathematics Research Centre, Coventry
University, Coventry, CV1 5FB, United Kingdom}
\address{$^2$ Department of Chemical Engineering, Imperial College London, London SW7 2AZ, United Kingdom}

\begin{abstract}
Conformations of a single-component bottle-brush polymer with a
fully flexible backbone under poor solvent conditions are studied
by molecular-dynamics simulations, using a coarse-grained
bead-spring model with side chains of up to $N=40$ effective
monomers. By variation of the solvent quality and the grafting
density $\sigma$ with which side chains are grafted onto the
flexible backbone, we study for backbone lengths of up to
$N_b=100$ the crossover from the brush/coil regime to the dense
collapsed state. At lower temperatures, where collapsed chains
with a constant monomer density are observed, the choice of the
above parameters does not play any role and it is the total number
of monomers that defines the dimensions of the chains.
Furthermore, bottle-brush polymers with longer side chains possess
higher spherical symmetry compared to chains with lower side-chain
lengths in contrast to what one may intuitively expect, as the
stretching of the side chains is less important than the increase
of their length. At higher temperatures, always below the Theta
($\Theta$) temperature, coil-like configurations, similar to a
single polymer chain, or brush-like configurations, similar to a
homogeneous cylindrical bottle-brush polymer with a rigid
backbone, are observed, depending on the choice of the particular
parameters $N$ and $\sigma$. In the crossover regime between the
collapsed state (globule) and the coil/brush regime the
acylindricity increases, whereas for temperatures outside of this
range, bottle-brush polymers maintain a highly cylindrical
symmetry in all configurational states.
\end{abstract}

\pacs{02.70Ns, 64.75.Jk, 82.35.Jk}

\submitto{\JPCM}

\maketitle

\section{Introduction}
\label{introduction}

Macromolecules with comb-like architectures, where linear or
branched side chains~\cite{Maleki2011} are grafted regularly or
randomly onto a backbone chain have recently found much interest
(for example, see the
reviews~\cite{Zhang2005,Subbotin2007,Sheiko2008,Potemkin2009,Binder2012,
Walther2013} and references therein). The interplay between steric
repulsion of the side-chain monomers and effective attractive
interactions that can be controlled by tuning the quality of the
solvent, leads to intricate spatial self-organization of these
molecular brushes. Such bottle-brush polymers can be either
flexible or stiff, depending, not only on the chemical character
of the backbone (which varies from a flexible macromolecule up to
a stiff nanorod, e.g., a carbon
nanotube~\cite{Meyyappan2004,Thomassin2010}), but, also, on the
chain length $N$ and the grafting density $\sigma$ of the side
chains. For example, by increasing $N$ and $\sigma$, stiffening of
intrinsically flexible backbones is induced, see,
e.g.,~\cite{Subbotin2007,Sheiko2008,
Rathgeber2005,Zhang2006,Hsu2010,Hsu2010b,Theodorakis2011}.
Moreover, conformations of such complex macromolecular objects can
be sensitive to external stimuli (electric fields, light, or
simply changes in the $p$H value of the solution). As a result,
various relevant applications have been
envisaged~\cite{Zhang2005,Sheiko2008}. In this work we do not
follow this aspect; rather we focus our interest on the
theoretical understanding of the structure formation of these
molecular brushes by exploring the statistical mechanics of a
coarse-grained model via molecular-dynamics (MD) simulations.

The study of the structure properties of bottle-brush polymers is
a challenging task due to the multitude of length scales
characterizing their
structure~\cite{Zhang2005,Rathgeber2005,Zhang2006,Hsu2010,Feuz2007}.
However, computer simulations have been very useful to validate
approximations needed to interpret experimental
data~\cite{Rathgeber2005,Hsu2010,Lee2008,Knaapilo2008,Klein2009,Shiokawa1999,Subbotin2000,Chang2009,Elli2004,
Connolly2005,Yethiraj2006,Hsu2007,Hsu2008}. Since some of the
suggested applications require to consider bottle-brush polymers
under poor solvent conditions, first steps to consider this case
by theory~\cite{Sheiko2004} and
simulations~\cite{Khalatur2000,Theodorakis2009,Theodorakis2010}
have been taken. In particular, for bottle-brush polymers with a
single type of side chains and rigid backbone, interesting
\textit{pearl-necklace} structures have been theoretically
predicted~\cite{Sheiko2004} and validated by numerical
simulations~\cite{Theodorakis2009,Theodorakis2010}. These studies
were also extended to bottle brushes with two different types of
side chains grafted alternately onto the
backbone~\cite{Theodorakis2011c, Theodorakis2011d,
Theodorakis2010b}, where such pearl-necklace structures were
observed at intermediate grafting densities. On the other hand,
Janus-type configurations were seen at high grafting densities.
The main hypothesis of these works was that the backbone is
strictly rigid falling in the case of a quasi-one-dimensional
system, disregarding in this way any effect attributed to the
flexibility of the backbone. The other extreme would be that the
backbone is considered fully flexible, as the side chains of the
brush. The latter case has only been studied for $\Theta$ and good
solvent conditions by computer
simulations~\cite{Theodorakis2011,Theodorakis2012} and it was
restricted to short chains, as the molecular weight in bottle
brushes increases very fast with increasing the grafting density
the backbone length and the length of the side chains. On the
other hand, long chains are usually considered in theoretical
work. As a result, a direct comparison with theoretical
predictions is a delicate matter. In the present study, we
consider for the first time the case of single-component
bottle-brush polymers with a fully flexible backbone under poor
solvent conditions. By varying the grafting density, the solvent
quality, and the backbone and side-chain length, we attempt to
give a complete picture of the self-organization properties of
such macromolecules in a poor solvent.

The rest of the paper is organized as follows: In the next section
we give a brief theoretical background; in section~\ref{model}, we
present the model and the simulation methods. Then,
section~\ref{results} defines and discusses the properties we
monitor in our MD simulations. We close this manuscript in
section~\ref{conclusions}, where we give a summary of our
conclusions and an outline of future work.

\section{Theoretical background}
\label{theoretical_background}

For bottle-brush polymers with a flexible backbone, no sharp phase
boundaries~\cite{Sheiko2004} are expected to occur as it has been
shown by MD simulations for bottle-brush polymers with a stiff
backbone~\cite{Theodorakis2009,Theodorakis2010}. Thus, we shall
not describe any detailed calculations on how to derive phase
boundaries here, but only give (background) qualitative arguments
to understand the phenomena observed in our simulations. The
cross-sectional linear dimension of a side chain of length $N$ in
the directions perpendicular to a rigid backbone and under good
solvent conditions is expressed as a scaling
relation~\cite{Hsu2007,Sheiko2004,
deGennes1979,Birshtein1984,Witten1986,Birshtein1987,Ligoure1990,Ball1991,Rouault1996,Saariaho1997}
\begin{equation}
\label{Eq1Sec2a} R_{cs}=N^{\nu}f(N^{\nu}\sigma) \rightarrow
\sigma^{(1-\nu)/(1+\nu)}N^{2\nu/(1+\nu)}.
\end{equation}
$\nu \approx 0.588$ is the characteristic exponent of a polymer
chain under good solvent conditions~\cite{LeGuillou1980}, $\sigma$
the grafting density, and $f$ is a scaling function which
describes the crossover from individual polymer \textit{mushrooms}
grafted onto the backbone of the bottle-brush with grafting
density $\sigma$. As usually, pre-factors of order of unity are
disregarded throughout. In this case, it has been
shown~\cite{Sheiko2004, Theodorakis2010b} that the densely grafted
side chains are stretched in the radial directions. For individual
polymer mushrooms the linear dimension $R_z$ in the direction
along the backbone also scales as $R_z \propto N^{\nu}$. Thus, the
quantity $N^{\nu}\sigma$ defines the distance that neighboring
side chains along the backbone start to interact, where the
distance between grafting sites is $\sigma^{-1}$. Then, excluded
volume interactions among the monomers of neighboring
\textit{mushrooms} start to cause chain stretching in radial
direction. However, we should note here the power-law behavior
$R_{cs} \propto N^{2\nu/(1+\nu)}$ can only be reached for very
long chains, for which there also exists a significant regime
where the radial monomer density profile decays with a related
power
law~\cite{Hsu2007,Birshtein1984,Witten1986,Birshtein1987,Ligoure1990,Ball1991,Rouault1996,Saariaho1997}
\begin{equation}
\label{Eq2Sec2a} \rho(r) \propto \left [ r/ \sigma \right
]^{-(3\nu-1)/2\nu} \approx \left [r/\sigma \right ]^{-0.65},
\qquad \alpha \ll r \leq R_{cs},
\end{equation}
$\alpha$ being the distance between (effective) monomers along a
side chain. As discussed in~\cite{Hsu2007}, for most cases of
practical interest, one neither reaches the regime where
equations~(\ref{Eq1Sec2a}) and (\ref{Eq2Sec2a}) hold, nor are
these equations suitable for chain lengths accessible to
experiments~\cite{Walther2013}.

Bottle-brush polymers under poor solvent conditions can be
described with similar arguments to that of a collapsed polymer
chain (see figure~\ref{fig1}). For a solution near the $\Theta$
temperature, scaling theories predict that the (gyration) radius
$R_c$ of the chain varies as~\cite{deGennes1979}
\begin{equation}
\label{Eq1Sec2} R_c=N^{1/2}F(N^{1/2}\tau), \qquad \tau=1-T/
\Theta,
\end{equation}
where the scaling function $F(Z)$ has the following limits:
\begin{eqnarray}
\label{Eq2Sec2}
F(Z \rightarrow \infty) \propto Z^{-1/3},\;\; R_c \propto \tau^{-1/3} N^{1/3},\;\; {\rm as} \;\; N \rightarrow \infty\nonumber \\
F(Z=0)={\rm const.}\nonumber\\
F(Z \rightarrow - \infty) \propto Z^{2\nu-1},\;\; R_c \propto
-\tau^{2\nu-1}N^{\nu},\;\; {\rm as} \;\; N \rightarrow \infty.
\end{eqnarray}
Here, $\nu=0.588$~\cite{LeGuillou1980} is the well-known
exponent~\cite{deGennes1979}, and the limits $N \rightarrow
\infty$ and $\tau \rightarrow 0$ are such that $|Z| \le \infty$.
As usually done, we ignore both logarithmic corrections, that are
expected to occur at the $\Theta$
point~\cite{desCloizeaux1990,Schaefer1999}, and corrections to
scaling that actually do become important for not very large
$N$~\cite{Theodorakis2011}.

One may often attempt to interpret scaling theories as a
comparison of lengths~\cite{deGennes1979}. The size of a polymer
coil in a $\Theta$ solvent is described by the scaling law
$R_c=\alpha N^{1/2}$, where $\alpha$ is the size of the monomeric
unit. Thus, the relation $Z=N^{1/2}\tau$ can also be written as
$Z=R_c/\xi$, $\xi=\alpha/\tau$, where the latter is the size of a
\textit{thermal blob}~\cite{deGennes1979,Rabin1994,Halperin1991}.
For collapsed globules, slightly below the $\Theta$ point
[equation~(\ref{Eq2Sec2})], the structure of the coil can be
described as a dense packing of blobs, each one of size
$\xi=\sqrt{n} \alpha$, where $n$ is the number of monomers per
blob. Then, the number of these blobs is
$n_b=N/n=N/(\xi/\alpha)^{2}$. The density inside this coil should
be essentially constant (figure~\ref{fig2}) and is described by
the following equation
\begin{equation}
\label{Eq4Sec2} \rho=n/\xi^{3}=1/(\alpha^2 \xi)=\tau/\alpha^3.
\end{equation}
The surface region of the globule is just the outer shell of the
blobs, of thickness $\xi$, giving rise to a surface free energy
(surface tension) $\gamma=n_{b}^{s}/R_c^2$, $n_b^s=n_b^{2/3}$
being the number of blobs in this outer shell (remember that the
free-energy cost of one blob is $k_B T$). Using
equation~(\ref{Eq2Sec2}) and omitting all geometric pre-factors
one end with
\begin{equation}
\label{Eq5Sec2} \gamma= \left ( \frac{n_b}{R^3_c} \right
)^{2/3}=\tau^2\alpha^{-2}.
\end{equation}
Equations~(\ref{Eq4Sec2}) and (\ref{Eq5Sec2}) characterize
isolated free chains and individually collapsed side chains onto
the rigid backbone of a bottle-brush, as long as the
$h=\sigma^{-1}$ is large enough, so that there is no free-energy
gain when neighboring collapsed globules coalesce. Such relations
can also describe bottle-brush polymers with a flexible backbone
in the collapsed state (see figure~\ref{fig2}). Thus, the total surface free energy of a
collapsed bottle-brush polymer with a flexible backbone is of the
order of $R_c^2\gamma=N^{2/3}\tau^{4/3}=(N\tau^2)^{2/3}$.

For rigid backbones, when the distance $h$ and the radius $R_c$ of
the collapsed chain [as described by equation~(\ref{Eq2Sec2})]
become comparable, the system can minimize its free-energy cost if
two, or more neighboring globular chains,
coalesce~\cite{Sheiko2004,Theodorakis2009,Theodorakis2010,
Theodorakis2010b}. If we assume that the number of thermal blobs
remains constant, the bulk free energy decides on the stability of
the different structures. The object formed by coalescence of two
collapsed coils that are grafted at a distance $h$, clearly will
result in an elongated object of length $L$ along the backbone and
radius $R$ in the plane perpendicular to the backbone, with $R \le
L$. The formation of such structures results in a free-energy
penalty of the order of $\tau$~\cite{Sheiko2004,Halperin1991}.
Thus, the free-energy cost of a cluster formed by $m$ neighboring
chains grafted within a distance $L=mh$ is~\cite{Sheiko2004}
\begin{equation}
\label{Eq6Sec2} \Delta F_1 (m) / k_B T = (mN\tau^2)^{2/3} -
m(N\tau^2)^{2/3}+m^2(h/\alpha)\tau.
\end{equation}
The total number of monomers in a cluster formed from $m$ chains
is $mN$, and hence the total surface free-energy cost of such an
object is $(mN\tau^2)^{2/3}$, whereas  the total surface
free-energy cost of $m$ separated globules, that contain $N$
monomers each, is $m(N\tau^2)^{2/3}$. The last term $m^2h\tau$ is
the total elastic stretching energy and scales quadratically in
$m$, as expected. Such an analysis is relevant for bottle brushes
with a rigid backbone, whereas for bottle-brush polymers with a
flexible backbone, only two regimes of rather homogenous
structures occur (see figure~\ref{fig1}): the collapsed chain
state for temperatures further from the $\Theta$ regime, and the
coil/brush regime, for temperatures close to the $\Theta$
temperature. Therefore, \textit{pearl-necklace}-like structures
are fully attributed to the backbone
stiffness~\cite{Theodorakis2009}.

\section{Model and simulation methods}
\label{model}

For bottle-brush polymers with flexible backbones under poor
solvent conditions~\cite{Theodorakis2009,Theodorakis2010} we
describe both the backbone chain and the side chains by a
bead-spring model~\cite{Grest1986,Murat1989,Binder1995}, where all
beads interact with a truncated and shifted Lennard-Jones (LJ)
potential $U_{\rm LJ}(r)$ and nearest neighbors, bonded together
along a chain, also experience the finitely extensible nonlinear
elastic (FENE) potential $U_{\rm FENE}(r)$, $r$ being the distance
between the beads. Thus,
\begin{equation}
\label{Eq1} U_{\rm LJ} (r)=4 \epsilon_{\rm LJ}\left[ \left(
\frac{\sigma _{\rm LJ}}{r}\right)^{12}-\left(\frac{\sigma _{\rm
LJ}}{r}\right)^6 \right] +{\rm C}, \qquad r \leq r_{c},
\end{equation}
while $U_{\rm LJ}(r>r_c)=0$, and $r_c=2.5 \sigma_{\rm LJ}$. The
constant $\rm C$ is defined such that $U_{\rm LJ}(r=r_c)=0$ is
continuous at the cut-off. Henceforth, units are chosen such that
$\varepsilon_{\rm LJ}=1$, $\sigma_{\rm LJ}=1$, the Boltzmann
constant $k_B=1$, and also the mass $m_{\rm LJ}$ of all beads is
chosen to be unity. The potential of equation~(\ref{Eq1}) acts
between any pair of beads, irrespective of whether they are bonded
or not. For bonded beads also the potential $U_{\rm FENE}(r)$
acts, where
\begin{equation}\label{Eq2}
 U_{\rm FENE}(r)=-\frac{1}{2} k r_{0}^{2}\ln\left[1-\left(\frac{r}{r_{0}}\right)^{2}\right]
\qquad 0<r\leq r_{0},
\end{equation}
$r_{0}=1.5$, $k=30$, and $U_{\rm FENE}(r)=\infty$ outside the
range written in equation~(\ref{Eq2}). Hence $r_0$ is the maximal
distance that bonded beads can take. Note that we did not include
any explicit solvent particles; solvent-mediated interactions and
solvent quality are only indirectly simulated by varying the
temperature of the system.

In our model there is no difference in interactions, irrespective
of whether the considered beads are effective monomers of the
backbone or side chains. This implies that, the polymer forming
the backbone is either chemically identical to the polymers that
are tethered as side chains to the backbone, or at least on
coarse-grained length scales, as considered here, the backbone and
side chain polymers are no longer distinct. There is also no
difference between the bond linking the first monomer of a side
chain to a monomer of the backbone and bonds between any other
pairs of bonded monomers. Of course, our study does not address
any effects due to a particular chemistry relating to the
synthesis of these bottle-brush polymers, but, as usually
done~\cite{Binder1995}, we address universal features of the
conformational properties of these macromolecules.

There is one important distinction comparing to our previous
work~\cite{Theodorakis2009} on bottle-brush polymers with rigid
backbones: in~\cite{Theodorakis2009} the backbone was taken,
following Murat and Grest~\cite{Murat1989}, as an infinitely thin
straight line in continuous space, thus allowing arbitrary values
of the distances between neighboring grafting sites, and hence the
grafting density $\sigma$ could be continuously varied. For the
present model, where we disregard any possible quenched disorder
resulting from the grafting process, of course, the grafting
density $\sigma$ is quantized: we denote here by $\sigma=1.0$ the
case that every backbone monomer carries a side chain,
$\sigma=0.5$ means that every second backbone monomer has a side
chain, etc. Chain lengths of side chains were chosen as
$N=4,10,20$, and $N=40$, while backbone chain lengths were chosen
as $N_b=50$ and $N_b=100$, respectively. It is obvious that for
such short side-chain lengths and under poor solvent conditions
any interpretation of characteristic lengths is a delicate matter
for specified range of rather small values of $N$. However, we
should emphasize that our range of $N$ nicely corresponds to the
range available in
experiments~\cite{Zhang2005,Sheiko2008,Rathgeber2005,Zhang2006,Hsu2010,Lecommandoux2002}
and that the simulation of longer chains becomes an extremely
cumbersome task, as relaxation times are increasing exponentially.
Already for bottle-brush polymers under $\Theta$ and good solvent
conditions equilibration of bottle-brush polymers with long
backbones~\cite{Theodorakis2011} (e.g., $N_b=200$) required an
enormous effort. We do not attempt to reach such backbone lengths
here.

Unfortunately, for the model defined in equations~(\ref{Eq1}) and
(\ref{Eq2}), the $\Theta$ temperature is known only rather
roughly~\cite{Grest1993}, namely, $\Theta \approx 3.0$. Being
interested in $T \leq \Theta$, we have attempted to study the
temperature range $1.5 \leq T \leq 3.0$. Note however that,
equilibration of collapsed chains is rather difficult, as will be
discussed below, besides the particular problems encountered due
to the multitude of length scales in a bottle-brush macromolecule.
In our simulations, the temperature is controlled by the Langevin
thermostat, following previous
work~\cite{Binder1995,Theodorakis2011b,Theodorakis2011e,
Theodorakis2011f, Theodorakis2012b,Murat1991}. The equation of
motion for the coordinates $\{r_i(t)\}$ of the beads
\begin{equation}\label{Eq3}
 m\frac{d^{2}\textbf{r}_{i}}{dt^{2}}=-\nabla U_{i}-
\gamma \frac{d\textbf{r}_{i}}{dt}+\Gamma_{i} (t)
\end{equation}
is numerically integrated using the GROMACS
package~\cite{Berendsen1995,Lindahl2001}. In equation~(\ref{Eq3}),
$t$ denotes the time, $U_{i}$ is the total potential acting on the
$i$-th bead, $\gamma$ is the friction coefficient, and
${\Gamma}_i(t)$ is the random force. As it is well-known, $\gamma$
and $\Gamma$ are related via the usual fluctuation-dissipation
relation
\begin{equation}\label{Eq4}
<\Gamma_{i}(t)\cdot
\Gamma_{j}(t^{'})>=6k_{B}T\gamma\delta_{ij}\delta(t-t^{'}).
\end{equation}
Following references~\cite{Binder1995,Grest1993,Murat1991}, the
friction coefficient was chosen as $\gamma=0.5$.
Equation~(\ref{Eq3}) was integrated with the leap-frog
algorithm~\cite{vanGunsteren1988} using an integration time step
of $\Delta t=0.006 \tau$, where the MD time unit is $\tau=(m_{\rm
LJ}\sigma_{\rm LJ}^{2}/\epsilon_{\rm LJ})^{1/2}=1$.

Firstly, the system was equilibrated at a temperature $T=3.0$
using simulations extending over a time range of $30\times 10^6
\tau$. To gather proper statistics, we used $500$ independent
configurations at this high temperature, as initial configurations
for slow cooling runs, where the temperature was lowered in steps
of $0.1$, and the system was simulated, at each temperature, for a
time of $5 \times 10^6 \tau$. The final configuration of each
(higher) temperature was used as starting configuration for the
next (lower) temperature. In this way, one can safely generate, at
low temperatures and intermediate values of grafting densities,
statistically independent configurations. Let us point out here
that, standard MD simulations would typically not sample phase
space adequately at this low-temperature regime. However, this
problem was surpassed here by carrying out this large number of
independent slow cooling runs (typically around $500$ runs). The
statistical accuracy of our results was also checked by varying on
the same footing the cooling speed and length of runs. For
temperatures above $T=2.0$ correlations are rather insignificant,
but for temperatures below $T=2.0$ this slow cooling methodology
is indispensable in order to get reliable results for our chosen
range of parameters.

\section{Results}
\label{results}

\begin{figure*}
\begin{center}
\rotatebox{0}{\resizebox{!}{0.60\columnwidth}{%
  \includegraphics{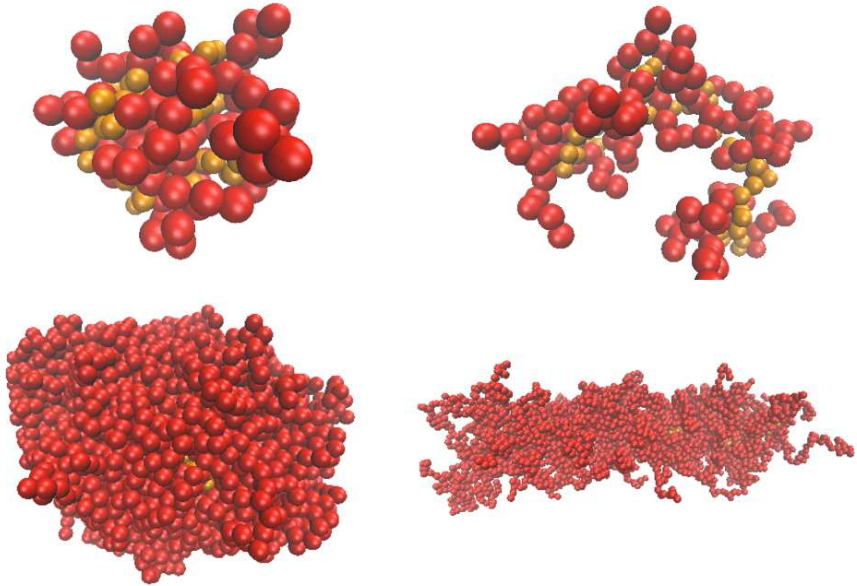}
}}
\end{center}
\caption{\label{fig1}(Color online) Characteristic snapshots of
bottle-brush polymers with a flexible backbone. In the left part
snapshots are taken at temperature $T=1.5$, whereas on the right
part at $T=3.0$. Upper part refers to a bottle-brush with
parameters $\sigma=0.5$, $N_b=50$, and $N=5$, whereas in the lower
part we consider $\sigma=1.0$, $N_b=100$, $N=40$. At low
temperatures, chains are in the collapsed state (globule), similar
to a single polymer chain, whereas at the higher temperature we
distinguish between a coil-like structure and an almost
homogeneous cylindrical brush similar to a homogeneous cylindrical
brush with a rigid backbone.}
\end{figure*}

In the case of a single-component bottle-brush polymer with a
strictly rigid backbone and at low grafting densities, side chains
collapse individually onto the backbone. When the grafting
density, or the side-chain length increases, chains start forming
clusters which consist of more than one side chain. Further
increase of the grafting density (or very long side chains) to the
so-called \textit{brush} regime, leads to an homogenously
cylindrical structure~\cite{Theodorakis2009,Theodorakis2010}. When
we study the case of a single-component bottle-brush polymer with
a fully flexible backbone under poor solvent conditions, we can
basically distinguish two regimes: At low temperatures, or poor
solvent conditions, relatively far from the $\Theta$ point, the
chain obtains an almost spherical structure, like those in the
left part of figure~\ref{fig1} for either lower or higher grafting
density, shorter or longer backbone, and shorter or longer side
chains (we shall discuss the shape of this spherical form later in
this section). On the other hand, at temperatures close to the
$\Theta$ temperature, gaussian-like chains are formed for very low
grafting density and short side chains or almost homogenous
cylindrical brushes as the grafting density and side-chain length
increase, as clearly shown in the right part of the same figure.

\begin{figure*}
\begin{center}
\rotatebox{270}{\resizebox{!}{0.50\columnwidth}{%
  \includegraphics{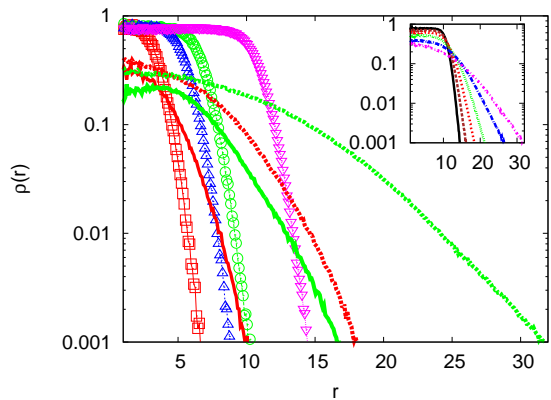}
}}
\end{center}
\caption{\label{fig2}(Color online) Density profiles as a function
of the distance from the center of mass of a bottle-brush at
$T=1.5$ (open symbols) and $T=3.0$ (lines). Squares or continuous
red line refer to $\sigma=0.5$, $N_b=50$, and $N=5$, circles or
the dashed red line to $\sigma=0.5$, $N_b=50$, and $N=40$,
triangles or the continuous green line to $\sigma=1.0$, $N_b=100$,
and $N=5$, and upside-down triangles or the dashed green line to
$\sigma=1.0$, $N_b=100$, and $N=40$. The inset illustrates the
density profile for the case $\sigma=1.0$, $N_b=100$, and $N=40$
as a function of temperature. The continuous black line
corresponds to $T=1.5$ and temperatures up to $T=3.0$ are shown
with step $\Delta T=0.3$. No signs of sharp phase transition
between the collapsed state and the brush regime are observed.}
\end{figure*}

The first quantity we have computed is the overall monomer
density, mathematically expressed by the following formula
\begin{equation}\label{eq:6}
\rho(|\vec{r}|)= \langle  \sum \limits_{i=1}^{nN} \delta(
\vec{r}-\vec{r}_{c}-\vec{r}_{i}) \rangle,
\end{equation}
where $\delta(\vec{x})$ is the Dirac delta function, $\vec{r}_c$
the position of the center of mass of all monomers that belong to
the chain, and $\vec{r}_i$ the positions of all monomers,
irrespective of whether they belong to the backbone or not. The
angle brackets denote an average over all conformations, as usual.
Figure~\ref{fig2} shows two characteristic examples of the overall
monomer density. At temperature $T=1.5$, it is clear that the
bottle-brush polymer, close to its center of mass, has a constant
density near to the melt polymer density, as expected. Then, the
density profile decays far from its center of mass and the
starting point of this decay depends only on the total number of
beads of the bottle-brush macromolecule. Therefore, any effects
attributed to the specific architecture of bottle-brush polymers
are not coming onto the surface at such low temperatures. We will
later see what is the range of temperature that corresponds to the
collapsed state. Furthermore, this density decay does not reveal
any significant behavior that could be attributed to the specific
macromolecular architecture. On the other hand, the effect of
architecture is more noticeable at temperatures close to the
$\Theta$ temperature, where the increase of the grafting density
$\sigma$ and the chain length $N$ results in a more extended
density profile with distance $r$. Furthermore, the gradual
lowering of the temperature from the brush regime to the collapsed
chain regime (inset of figure~\ref{fig2}) does not hint for any
sharp transition; we observe a smooth variation of the density
profile as the temperature is decreased. The same behavior is
observed for all the bottle brushes of this study.

Following the definition of standard textbooks, aspects related to the
persistence length of the polymer chains may be discussed,
where the persistence length is defined from the decay of bond
orientational correlations along the chain~\cite{Theodorakis2011}. However, one should be
careful when attempts to estimate the persistence length of polymer chains,
as for solvent conditions other than $\Theta$, different
definitions account for different values of the persistent length~\cite{Theodorakis2011}.
Defining the bond vectors $\vec{\alpha}_{i}$ in terms of the
monomer positions $\vec{r}_{i}$ as
$\vec{\alpha}_{i}=\vec{r}_{i+1}-\vec{r}_{i}$, $i=1,...,nN-1$, the
bond orientational correlation is given by
\begin{equation}
\label{eq:5} \langle cos \Theta(s) \rangle= l_{b}^{-2}
\frac{1}{N_{b}-1-s}  \sum_{i=1}^{N_ {b}-1-s} \langle
\vec{\alpha}_{i} \cdot \vec{\alpha}_{i+s} \rangle.
\end{equation}
Note that $\vec{\alpha}_{i}^2=l_{b}^{2}$, with $l_{b}$ being the
average bond length. For this model $l_{b} \approx 1$ and hence
$\langle cos \Theta(0) \rangle \approx 1$. Considering the limit
$nN \rightarrow \infty$ and assuming Gaussian chain statistics,
one obtains an exponential decay, since then $\langle cos\Theta(s)
\rangle= \langle cos\Theta(1) \rangle^{s}= \exp[\ln\langle
cos\Theta(1) \rangle]$, and thus
\begin{equation}
\label{eq:6} \langle cos\Theta(s) \rangle= \exp[-s/l_{p}],
l_{p}^{-1}=-\ln\langle cos\Theta(1) \rangle.
\end{equation}
For chains at the $\Theta$ point~\cite{Shirvanyants2008} or in
melts~\cite{Beckrich2007} one has
\begin{equation}
\label{eq:7} \langle cos\Theta(s) \rangle \propto s^{-3/2},
s\rightarrow \infty.
\end{equation}
By measuring this property we have verified that in the collapsed
states of figure~\ref{fig1} correlations along the backbone die
out very quickly and within the range of $2-4$ monomers, depending
on the choice of parameters. Of course, for the case $\sigma=1.0$
and $N=40$ some more correlation is observed, which, however, does
not justify any further analysis of these data. Also, we were not
able to detect any kind of correlation patterns similarly to the
case of multi-block copolymers in a poor
solvent~\cite{Theodorakis2012b}.

\begin{figure*}
\begin{center}
\rotatebox{270}{\resizebox{!}{0.50\columnwidth}{%
  \includegraphics{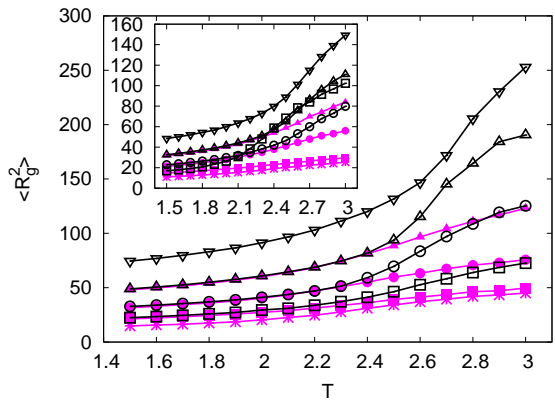}
}}
\end{center}
\caption{\label{fig3} (Color online) Mean-square gyration radius
for the whole bottle-brush polymer as a function of the
temperature. In the main part data refer to $\sigma=1.0$, whereas
the inset illustrates data for $\sigma=0.5$. Full symbols
correspond to $N_b=50$ and open symbols to $N_b=100$ for both
grafting densities. Curves correspond to $N=5,10,20$, and $N=40$,
starting from the bottom.}
\end{figure*}

Turning back our discussion to the properties related to the size
of the bottle-brush polymers, we calculated properties such as the
mean-square gyration radius of the whole bottle brush $\langle
R^2_g \rangle$, backbone $\langle R^2_{g,b} \rangle$, and side
chains $\langle R^2_{g,s} \rangle$, as well the end-to-end
distance of the backbone $\langle R^2_{e,b} \rangle$.
Figure~\ref{fig3} presents results for the mean-square gyration
radius of the whole macromolecule, as a function of the
temperature. This property confirms our conclusions from
figure~\ref{fig2} that bottle brushes with the same number of
monomers at low temperatures have the same dimensions in the
collapsed state, achieving a constant density close to the center
of mass and a similarly decay pattern far from the center. This
behavior persists over a higher range of temperatures, as the
total number of monomers increases, and provides a rough
estimation of the onset of the collapsed state as the temperature
is lowered. This behavior is seen at both $\sigma=0.5$ and $1.0$
cases.

\begin{figure*}
\begin{center}
\rotatebox{270}{\resizebox{!}{0.50\columnwidth}{%
  \includegraphics{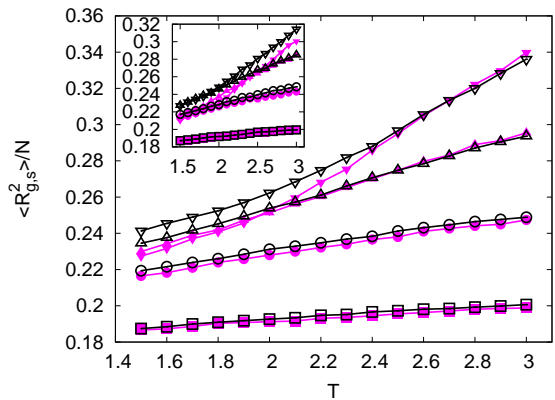}
}}
\end{center}
\caption{\label{fig4} (Color online) Mean-square gyration radius
of the side chains divided by the side-chain length $N$ as a
function of temperature for different cases. Symbols are chosen in
the same way as in figure~\ref{fig3}.}
\end{figure*}

\begin{figure*}
\begin{center}
\rotatebox{270}{\resizebox{!}{0.50\columnwidth}{%
  \includegraphics{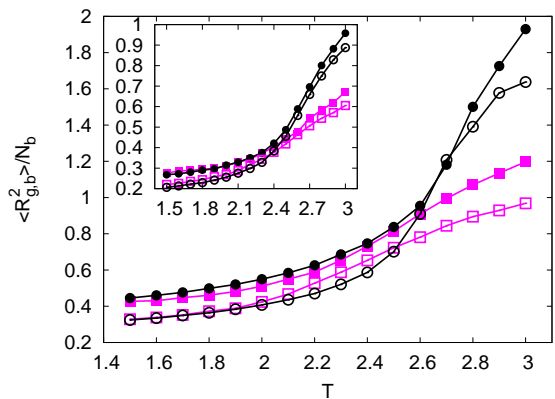}
}}
\end{center}
\caption{\label{fig5} (Color online) Mean-square gyration radius
of the bottle-brush backbone normalized by the backbone chain
length $N_b$ as a function of the temperature for $N=40$ (full
symbols) and $N=20$ (open symbols). We consider $\sigma=1.0$ (main
panel) and $\sigma=0.5$ (inset). Black lines and symbols
correspond to $N_b=100$, whereas the corresponding magenta colored
lines and symbols to $N_b=50$.}
\end{figure*}

\begin{figure*}
\begin{center}
\rotatebox{270}{\resizebox{!}{0.50\columnwidth}{%
  \includegraphics{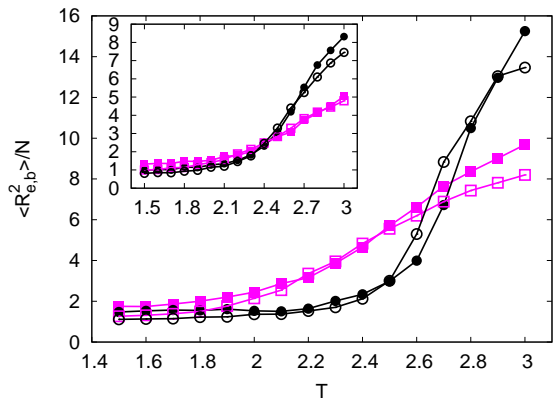}
}}
\end{center}
\caption{\label{fig6} (Color online) Same as in figure~\ref{fig5},
but the end-to-end distance of the bottle-brush backbone is
plotted as a function of the temperature.}
\end{figure*}

On the other hand, $\langle R^2_{g,s} \rangle$ exhibits a
dependence on the particular parameters of the bottle brush for
the whole range of temperatures, below the $\Theta$ temperature
(we refer to figure~\ref{fig4}). However, it is a subtle issue to
make any quantitative statement, as our analysis is restricted to
short side-chain lengths. Qualitatively, one may note that the
side chains and the backbone rearrange themselves in such a way at
low temperatures, that the overall density close to the center of
mass remains constant. This is what is seen if one combines the
data of figures~\ref{fig4} and \ref{fig5}. Furthermore, by
plotting the data for different chain lengths, such as the data of
figure~\ref{fig5} for the backbone dimensions normalized with
$N_b$, we get an estimate for an ``effective'' $\Theta$
temperature of the backbone. Apparently, the $\Theta$ temperature
of the model for a single polymer chain is different to that of
the bottle-brush backbone due to the presence of the side chains,
which introduce an overall stiffness on the macromolecule. For
$N=20$ this crossing of the curves for $N_b=50$ and $N_b=100$, may
suggest a $\Theta$ temperature while keeping constant $\sigma$ and
$N$. On the other hand, as $N$ increases to $40$, we obtain a
behavior that is similar to the case of a cylindrical bottle-brush
with a rigid backbone~\cite{Theodorakis2009}. That is an overlap
of the curves at low enough temperatures. A detailed discussion
about this feature can be found in~\cite{Theodorakis2009}. It
would be also interesting to obtain data for longer backbones, but
this is an enormously difficult task for bottle-brush
macromolecules under poor solvent conditions. Already under
$\Theta$ and good solvent conditions equilibration of such
macromolecules with long backbones has been proven extremely
challenging (e.g., $N_b=200$) due to the existence of the
so-called \textit{breathing modes}~\cite{Theodorakis2011}.
Measuring the end-to-end distance of the bottle-brush backbone,
figure~\ref{fig6}, reveals similar effects, but this quantity is
more reliable when one makes a comparison with the theoretical
predictions. Although our study is limited by the chain lengths,
we do not expect any qualitatively different behavior in the
long-chain limit.

\begin{figure*}
\begin{center}
\rotatebox{270}{\resizebox{!}{0.50\columnwidth}{%
  \includegraphics{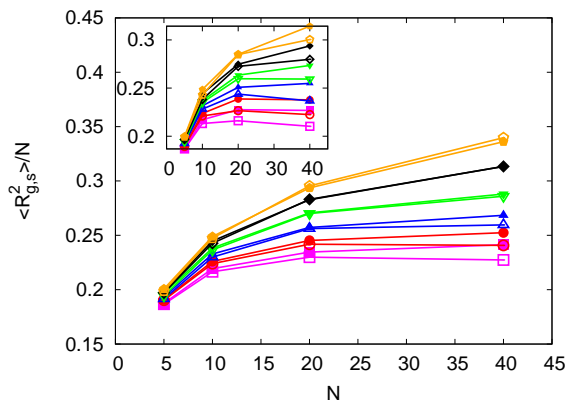}
}}
\end{center}
\caption{\label{fig7} (Color online) Mean-square gyration radius
as a function of the side-chain length $N$. Full symbols
correspond to $N_b=50$, whereas open symbols to $N_b=100$. Data
for different temperatures are shown, starting from below $T=1.5$,
up to $T=3.0$. In the main part of the figure $\sigma=1.0$. The
inset illustrates the corresponding data for the case
$\sigma=0.5$.}
\end{figure*}

Figure~\ref{fig7} presents results for the mean-square gyration
radius of the side chains as a function of the chain length $N$.
Whereas for higher temperatures one can observe the onset of a
scaling regime with $N$, as expected (see
also~\cite{Theodorakis2011}), at lower temperatures, and $N>20$,
there is only a weak dependence on the side-chain length $N$. This
is an indication that also the side chains themselves exhibit a
constant density regime close to their center of mass, similarly
to the case of the whole bottle-brush macromolecule. It is rather
naive and misleading to discuss any attempts of scaling, but we
have found good agreement with $N^{1/3}$, for our backbones of the
bottle-brush macromolecule.

An interesting aspect of our discussion is based on the study of
the shape of these collapsed objects at low enough
temperatures~\cite{Theodorakis2012, Theodorakis2012b}. We follow
the description of Theodorou and Suter~\cite{Theodorou1985} to
define the asphericity and acylindricity of a multi-block
copolymer chain. First, we define the gyration tensor as
\begin{equation}
\label{eq:8} \mathbf{S} = \frac{1}{N_{\xi}} \sum_{i=1}^{N_{\xi}}
\mathbf{s}_{i} \mathbf{s}_{i}^{T}= \overline{\mathbf{s}
\mathbf{s}^{T}}=
%\[
\left[ {\begin{array}{ccc}
  \overline{x^2} & \overline{xy} & \overline{xz} \\
  \overline{xy} & \overline{y^2} & \overline{yz} \\
  \overline{xz} & \overline{yz} & \overline{z^2} \\
 \end{array} } \right],
% \]
\end{equation}
where $\mathbf{s}_i=col(x_i,y_i,z_i)$ is the position vector of
each bead, which is considered with respect to the center of mass
of the beads $\sum_{i=1}^{N_{\xi}} \mathbf{s}^i=0$, and the
over-bars denote an average over all beads $N_{\xi}$. When the
gyration tensor of the whole chain is considered, then
$N_{\xi}=nN$. For the blocks $N_{\xi}=N$, the gyration tensor and
relevant properties are calculated for each block separately and
then an average over all blocks is taken, irrespective of their
type. The gyration tensor is symmetric with real eigenvalues and,
hence, a cartesian system that this tensor is diagonal can always
be found, i.e.,
\begin{equation}
\label{eq:9} \mathbf{S}=diag(\overline{X^2}, \overline{Y^2},
\overline{Z^2}),
\end{equation}
where the axes are also chosen in such a way that the diagonal
elements (eigenvalues of $\mathbf{S}$) $\overline{X^2}$,
$\overline{Y^2}$,  and $\overline{Z^2}$ are in a descending order
($\overline{X^2} \ge \overline{Y^2} \ge \overline{Z^2}$). These
eigenvalues are called the principal moments of the gyration
tensor. From the values of the principal moments, one defines
quantities such as the asphericity $b$
\begin{equation}
\label{eq:10} b= \overline{X^2} -1/2( \overline{Y^2} +
\overline{Z^2}).
\end{equation}
When the particle distribution is spherically symmetric or has a
tetrahedral or higher symmetry, then $b=0$. The acylindricity $c$
\begin{equation}
\label{eq:11} c=\overline{Y^2} - \overline{Z^2}
\end{equation}
is zero when the particle distribution is in sync with a
cylindrical symmetry. Therefore, the acylindricity and asphericity
are relevant quantities that would characterize the shape of our
bottle-brush polymers. These quantities are taken with respect to
$S$, to the sum of the eigenvalues, i.e. the square gyration
radius of the chain, which we have also calculated independently
on our original cartesian coordinates, in order to check our
results. Henceforth, this quantity will be absorbed in the
definition of asphericity and acylindricity and all results will
be normalized under this norm.

\begin{figure*}
\begin{center}
\rotatebox{270}{\resizebox{!}{0.50\columnwidth}{%
  \includegraphics{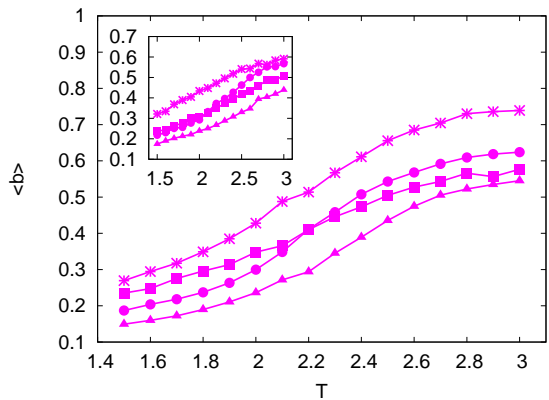}
}}
\end{center}
\caption{\label{fig8} (Color online) Normalized asphericity as a
function of the temperature. The main panel shows data for
$\sigma=1.0$, whereas $\sigma=0.5$ in the inset. The cases
$N_b=50$ and $N=5$ (stars), $N=10$ (squares), $N=20$ (circles),
and $N=40$ (triangles) are shown.}
\end{figure*}

\begin{figure*}
\begin{center}
\rotatebox{270}{\resizebox{!}{0.50\columnwidth}{%
  \includegraphics{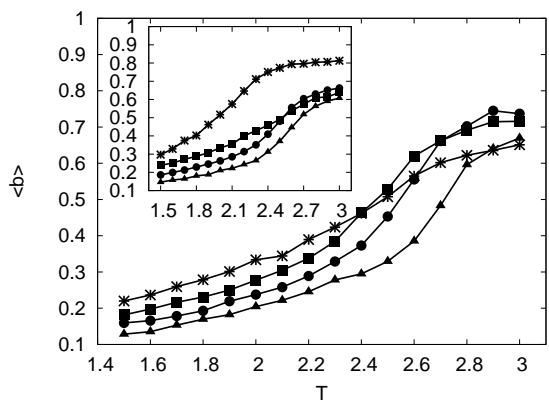}
}}
\end{center}
\caption{\label{fig9}Same as in figure~\ref{fig8}, but for
$N_b=100$.}
\end{figure*}

Assuming an ellipsoidal shape and based on the calculation of the
above eigenvalues of the gyration tensor, one can define an
effective volume expression~\cite{Hadizadeh2011}
\begin{equation}
\label{eq:12} V^{eff}=4\pi \sqrt[]{3} \prod_{i=1}^{3}
\sqrt[]{\lambda_{i}},
\end{equation}
where $\lambda_{1}= \overline{X^2}$, $\lambda_{2}=\overline{Y^2}$,
and $\lambda_{3}=\overline{Y^2}$, are the eigenvalues of the
radius of gyration tensor. Then, the effective radius of a sphere
with the same volume as this ellipsoid ($V^{eff}$) is given by the
geometrical mean of individual radii $R_{g}^{eff}= \sqrt[]{3}
\prod_{i=1}^{3} \sqrt[6]{\lambda_{i}}$. This can be compared to
the volume of an effective sphere defined by the gyration radius
$R_{g}=\sqrt[]{\sum^{i=1}_{3} \lambda_{i}}$. Such an analysis goes
clearly beyond the scope of the present study.

\begin{figure*}
\begin{center}
\rotatebox{270}{\resizebox{!}{0.50\columnwidth}{%
  \includegraphics{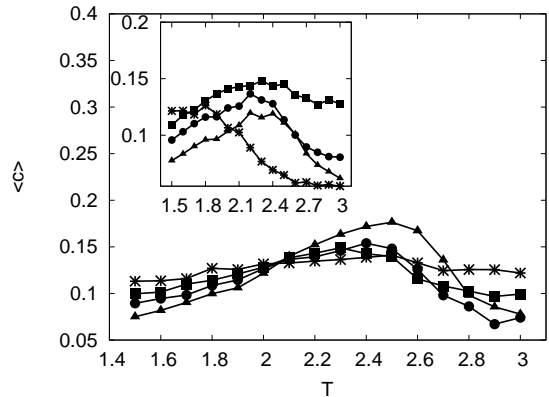}
}}
\end{center}
\caption{\label{fig10} Normalized acylindricity as a function of
the temperature. The main panel's data refer to grafting density
$\sigma=1.0$, whereas those of the inset to $\sigma=0.5$.
$N_b=100$ in both panels. The cases $N=5$ (stars), $N=10$
(squares), $N=20$ (circles), and $N=40$ (triangles) are shown.}
\end{figure*}

\begin{figure*}
\begin{center}
\rotatebox{270}{\resizebox{!}{0.50\columnwidth}{%
  \includegraphics{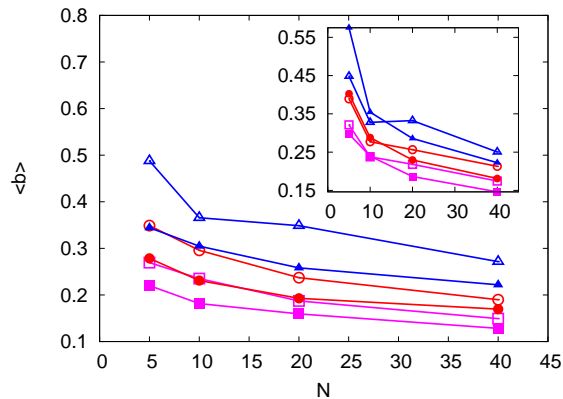}
}}
\end{center}
\caption{\label{fig11} (Color online) Normalized asphericity as a
function of the side-chain length $N$ for $\sigma=1.0$ and
$\sigma=0.5$ (inset). Open symbols refer to $N_b=50$, whereas full
symbols to $N_b=100$. Different temperatures are shown $T=1.5$
(squares), $T=1.8$ (circles), and $T=2.1$ (triangles).}
\end{figure*}

In figure~\ref{fig8} we show results for the asphericity as a
function of temperature for macromolecules with backbone length
$N_b=50$. At temperatures below $T=2.2$, where for all
bottle-brush cases collapsed brushes occur, we observe that chains
with longer side chains overall adopt a more spherical shape. This
means that the increase of the side chains stiffens the backbone
is not enough to change the overall picture of the macromolecule,
that is a more spherical shape for higher symmetry structure as
the side-chain length is increased. The results for backbone
length $N_b=100$ shown in figure~\ref{fig9} illustrate the same
behavior. Furthermore, as the temperature increases, there is a
significant change in the shape of the bottle brush from a
globular coil to an elongated object. Although we can see that we
depart from globular structure, we can say very little on whether
we obtain a coil-like structure or a homogeneous bottle brush
similar to a collapsed brush with a rigid
backbone~\cite{Theodorakis2009} from this quantity. Furthermore,
it is worth noting (figure~\ref{fig10}) that such non spherical
objects possess a cylindrical symmetry, even at low temperatures,
and this cylindrical symmetry is only distorted at intermediate
temperatures, i.e., within the crossover regime between collapsed
chains (globules) and coils or cylindrical brushes. This effect,
albeit weak and with a peak appearing at different temperatures
depending on the choice of parameters, is clearly shown in
figure~\ref{fig10}, where for temperatures in the range $2 < T <
2.6$ $\langle c \rangle$ a shallow maximum is observed. Finally,
regarding the dependence of the bottle-brush asphericity on the
length of the side chains $N$ in the collapsed state, we can
observe in figure~\ref{fig11} that the increase of the side-chain
length $N$ favors a more spherical symmetry for the bottle brush,
advancing in this way the backbone stiffening.

\section{Synopsis}
\label{conclusions}

In this study we have discussed the structural properties of
single-component bottle-brush polymers under poor solvent
conditions. We attempted to give a complete picture of the
structural properties of these macromolecules within the feasible
range of chain lengths, accessible to our simulations. For our
range of parameters, which is also similar to the range appearing
in
experiments~\cite{Zhang2005,Sheiko2008,Rathgeber2005,Zhang2006,Hsu2010,Lecommandoux2002},
we identified the different configurational states for a
bottle-brush polymer under poor solvent conditions. At low
temperatures, the bottle-brush chain collapses, forming
spherical-like objects with a constant density close to the center
of mass of the macromolecule, irrespective of the choice of
parameters $\sigma$, $N_b$, and $N$. Thus, the density profile
showed a dependency only on the total number of monomers. For
higher temperatures, there is a crossover from such collapsed
configurations to coil-like structures or cylindrical brushes
depending on the choice of the parameters $\sigma$, $N_b$, and
$N$. For small grafting densities, backbone, and chain lengths,
coil-like structures are favored, whereas, as the values of the
above properties increase, we smoothly obtain brushes with
monomers homogeneously distributed around the backbone, similar to
a bottle-brush polymer with a rigid backbone
\cite{Theodorakis2009}.

An interesting aspect regarding the shape of bottle-brush polymers
with a flexible backbone is that the increase of the side-chain
length overwhelms the stretching of the side chains, which is
caused by the steric repulsions between monomers, as the
side-chain length increases. For this reason, bottle brushes of
smaller $N$ have lower spherical or higher symmetry compared to
brushes with higher $N$, in contrast to what one may naively
expect. We also studied the acylindricity of bottle-brush
polymers. We found that the macromolecule maintains a highly
cylindrical symmetry for all these states. Only within the range
of temperatures corresponding to the crossover regime between the
globular (collapsed) state and coil/brush regime a weak deviation
from this behavior is observed. Overall, our results consist a
first attempt towards the study of bottle-brush polymers with a
flexible backbone and a first step in order to compare results
with the case of two-component bottle-brush polymers with a
flexible backbone under poor solvent conditions.

\end{document}